\documentclass[apjl]{emulateapj}
\usepackage[utf8]{inputenc}
\usepackage{graphicx}
\usepackage{comment}
\usepackage{natbib}
\usepackage{hyperref}
\usepackage{ulem,color}

\newcommand{\agal}{\ensuremath{a_{\rm Gal}}}
\newcommand{\aobs}{\ensuremath{a_{\rm LOS}^{\rm Obs}}}

\newcommand{\PBDOTobs}{\ensuremath{\dot{P}_b^{\rm Obs}}}
\newcommand{\PBDOTshk}{\ensuremath{\dot{P}_b^{\rm Shk}}}
\newcommand{\PBDOTgal}{\ensuremath{\dot{P}_b^{\rm Gal}}}
\newcommand{\PBDOTGR}{\ensuremath{\dot{P}_b^{\rm GR}}}
\newcommand{\Deltci}{\ensuremath{\Delta\rm t_{c,i}}}
\newcommand{\PSUAA}{Department of Astronomy \& Astrophysics, 525 Davey Laboratory, The Pennsylvania State University, University Park, PA, 16802, USA}
\newcommand{\PSUCEHW}{Center for Exoplanets and Habitable Worlds, 525 Davey Laboratory, The Pennsylvania State University, University Park, PA, 16802, USA}
\newcommand{\PSETI}{Penn State Extraterrestrial Intelligence Center, 525 Davey Laboratory, The Pennsylvania State University, University Park, PA, 16802, USA}

\begin{document}

\title{Eclipse timing the Milky Way's gravitational potential}

\author{Sukanya Chakrabarti\altaffilmark{1,2}, Daniel J. Stevens\altaffilmark{3,4}, Jason Wright\altaffilmark{3,4,5}, Roman R. Rafikov\altaffilmark{1,8}, Philip Chang\altaffilmark{7}, Thomas Beatty\altaffilmark{6}, Daniel Huber\altaffilmark{9}}


\altaffiltext{1}{Institute of Advanced Study, 1 Einstein Drive
Princeton, New Jersey
08540 USA; chakrabarti@ias.edu}
\altaffiltext{2}
{School of Physics and Astronomy, Rochester Institute of Technology, 84 Lomb Memorial Drive, Rochester, NY 14623}

\altaffiltext{3}{\PSUAA}
\altaffiltext{4}{\PSUCEHW}
\altaffiltext{5}{\PSETI}

\altaffiltext{6}
{Department of Astronomy and Steward Observatory, University of Arizona, Tucson, AZ 85721}
\altaffiltext{7}
{Department of Physics, University of Wisconsin-Milwaukee, 3135 North Maryland Avenue, Milwaukee, WI 53211}
\altaffiltext{8}
{Centre for Mathematical Sciences, Department of Applied Mathematics and Theoretical Physics, University of Cambridge, Wilberforce Road, Cambridge CB3 0WA, UK}
\altaffiltext{9}
{Institute for Astronomy, University of Hawai`i, 2680 Woodlawn Drive, Honolulu, HI 96822, USA}

\begin{abstract}
We show that a small, but \textit{measurable} shift in the eclipse mid-point time of eclipsing binary (EBs) stars of $\sim$ 0.1 seconds over a decade baseline can be used to directly measure the Galactic acceleration of stars in the Milky Way at $\sim$ kpc distances from the Sun.  We consider contributions to the period drift rate from dynamical mechanisms other than the Galaxy's gravitational field, and show that the Galactic acceleration can be reliably measured using a sample of $\textit{Kepler}$ EBs with orbital and stellar parameters from the literature.  Given the uncertainties on the formulation of tidal decay, our approach here is necessarily approximate, and the contribution from tidal decay is an upper limit assuming the stars are not tidally synchronized.   We also use simple analytic relations to search for well-timed sources in the \textit{Kepler} field, and find $\sim$ 70 additional detached EBs with low eccentricities that have estimated timing precision better than 1 second.   We illustrate the method with a prototypical, precisely timed EB using an archival \textit{Kepler} light curve and a modern synthetic \textit{HST} light curve (which provides a decade baseline). This novel method establishes a realistic possibility for obtaining fundamental Galactic parameters using eclipse timing to measure Galactic accelerations, along with other emerging new methods, including pulsar timing and extreme precision radial velocity observations.  This acceleration signal grows quadratically with time.  Therefore, given baselines established in the near-future for distant EBs, we can expect to measure the period drift in the future with space missions like \textit{JWST} and the \textit{Roman Space Telescope}.  
\end{abstract}

\section{Introduction}
Measurements of the accelerations of stars provide the most direct probe of the mass distributions (the stars and the dark matter) of galaxies.  For more than a century now, the fundamental parameters that describe the Milky Way (MW) have been determined using kinematic estimates of the accelerations of stars \citep{Oort1932,Kuijken_Gilmore1989,Bovy_Tremaine2012, McKee2015, Schutz2018} that live within its gravitational potential.  Recent advances in technology have led to the development of several different techniques to measure Galactic accelerations directly using extremely precise time-series measurements.

Extreme precision spectrographs that can achieve an instrumental precision of $\sim 10\,{\rm cm\,s}^{-1}$ \citep{Pepe2010,WrightRobertson} have opened up a new avenue to measure the Galactic acceleration directly \citep{Silverwood2019,Chakrabarti2020}. Analysis of ongoing pulsar timing observations have recently enabled a measurement of Galactic accelerations \citep{Chakrabarti2021} and constraints on the Galactic potential, including a measurement of the Oort limit, the local dark matter density, and the oblateness of the Galactic potential as traced by the pulsars.  
In this Letter, we develop a framework for using eclipsing binaries (EBs) to measure Galactic accelerations directly.

There is now a plethora of observed phenomena, in both the gas and the stellar disk, that indicates that our Galaxy has had a highly dynamic history \citep{LevineBlitzHeiles2006, ChakrabartiBlitz2009,ChakrabartiBlitz2011,Xu2015,helmi2018,antoja2018}.  This dynamic picture of the Galaxy has come into especially sharp focus with the advent of \textit{Gaia} data \citep{Gaia2016:main}. Analysis of interacting-MW simulations shows that there
are differences between the ``true'' density in these simulations and density estimates from the Jeans analysis (which assumes equilibrium) \citep{Haines2019}; this underlines the need for direct acceleration measurements that are based on time-series observations.  The acceleration profiles in interacting-MW simulations are highly asymmetric, in contrast to static potentials or isolated MW simulations \citep{Chakrabarti2020}.  For a galaxy with a dynamic history like the Milky Way, kinematic analyses based on snapshots of stars' positions and velocities do not fully capture the complexity of the Galactic mass distribution. 

In our earlier work \citep{Chakrabarti2021}, we used pulsar timing to measure accelerations directly and found differences at the level of a factor of $\sim$ 2 for the observed line-of-sight acceleration of the pulsars in our sample, compared to static potentials that are based on the Jeans analysis to estimate accelerations, which may be due to out-of-equilibrium effects.  The Oort limit we measured (to 3-sigma) is about 15 \% lower than that determined from the Jeans analysis \citep{McKee2015, Schutz2018}.  The oblateness of the potential traced by the pulsars is significantly closer to that of a disk rather than a halo. However, the pulsar sample is small (our earlier analysis included 14 binary millisecond pulsars that were timed sufficiently precisely such that we could extract the Galactic signal) and grows slowly, and so we are prompted to further explore the development of new direct acceleration techniques.

EBs have long been amenable to precise characterization, including $\leq 3\%$ measurements of their masses and radii \citep{Torres2010}. The advent of continuous, high-precision photometry from space telescopes such as \textit{Kepler} and \textit{TESS} has led to comparable-or-better levels of precision being achieved for an increasingly large sample of EBs (\citealt{Southworth2015,Southworth2021,Prsa2021}); in particular, \textit{Kepler}'s high photometric precision -- $\lesssim 500$ parts-per-million, or ppm, for its long-cadence observation of EBs at magnitudes $K \lesssim 15$ -- and long baseline permit measurements of EBs' eclipse times to sub-second precision (e.g. \citealt{ClarkCunningham:2019,Helminiak2019,Windemuth2019}). The change in the eclipse mid-point time due to the Galactic acceleration grows quadratically with time.  At $\sim$ kpc distances, we expect the eclipse mid-point time to have shifted by $\sim$ 0.1s in the decade between the \textit{Kepler} observations and today, i.e., to measure the Galactic signal, we require a eclipse timing precision of $\sim$ 0.1 s from both the archival \textit{Kepler} data and a light curve today. 

The outline of the Letter is as follows.  In \S 2, we review the various physical mechanisms that can change a binary's orbital period, including the general-relativistic (GR) precession of an eccentric orbit, tidal decay, tidally and rotationally induced quadrupole moments, and the acceleration exerted on the binary by planetary companions, as well as the acceleration induced by the Galactic potential.  Our goal here is to determine the part of the parameter space where contaminants to the Galactic signal are sufficiently minor that we can reliably measure the small shift in the eclipse mid-point time.  We also analyze sources from a recent compilation of EBs with custom-extracted lightcurves \citep{Windemuth2019} for which the orbital and stellar parameters were presented.  In \S 3, we present the expected timing precision for the set of sources from the \cite{Windemuth2019} paper for which contaminants to the Galactic signal should not be significant, as well as an additional 70 sources from the Kepler EB Catalog (\citealt{Prsa2011,Kirk2016}) for which we estimate sufficiently precise mid-eclipse times to enable a measurement of the Galactic acceleration today.  We also discuss a prototypical EB and calculate its simulated $\textit{HST}$ lightcurve and expected timing precision.  We conclude in \S 4. 

\section{Mechanisms that contribute to the period drift rate of eclipsing binaries} 
For an EB with binary orbital period $P_{b}$, various physical mechanisms can induce a change in the observed binary period over time, thus affecting the observed mid-eclipse time, $t_{c}$.  These include the contribution from the Galactic gravitational potential $\PBDOTgal$, the Shlovskii effect $\PBDOTshk$, the relativistic precession of an eccentric orbit $\PBDOTGR$; these first three effects also impact the measured time-rate of change of the binary period for pulsars, which we analyzed in our earlier work \citep{Chakrabarti2021}.  Additionally, circumbinary planets may affect the drift rate of the binary period, $\dot{P_{b}}^{\rm pl}$ 
; this last term can also affect pulsar timing, but one can place limits on possible planetary companions from existing pulsar timing data for even distant planetary companions, as in \citet{Kaplan2016}.  For stars that behave as fluid bodies there are several additional effects that are also important: tidal decay, $\dot{P_{b}}^{\rm tidal}$ and rotationally and tidally induced quadrupoles, $\dot{P_{b}}^{\rm quad/rot}$ and $\dot{P_{b}}^{\rm quad/tidal}$. As there are significant and well-known sources of uncertainties in the tidal decay formulation \citep{Ogilvie2014,Patra2020}, our approach here is necessarily approximate.  The sum of these various mechanisms will then lead to the observed time-rate of change of the binary period \PBDOTobs:

\begin{eqnarray}
\PBDOTobs &=& \PBDOTgal + \PBDOTshk + \PBDOTGR + \dot{P_{b}}^{\rm tidal} +  \dot{P_{b}}^{\rm quad/rot}\nonumber \\ 
 &&+  \dot{P_{b}}^{\rm quad/tidal} + \dot{P_{b}}^{\rm pl} .
\end{eqnarray}
\label{eq:pbdotobs}

The Galactic acceleration is $\agal$, which for simplicity we take to be a Gaussian centered at $\sim ~\rm 10~cm/s/decade$ for stars at $\sim$ kpc distances from the Sun \citep{Chakrabarti2020,Chakrabarti2021}.  Thus, we write the Galactic contribution to the time-rate of change of the binary period as:
\begin{equation}
\dot{P}_{b}^{\rm Gal} = \frac{\agal}{c}P_{b}.
\end{equation}
We summarize below the additional contributions to \PBDOTobs, and we follow closely the notation in \cite{Rafikov2009} in which $n = 2\pi /P$ is the orbital frequency, $\mu$ the proper motion in the plane of the sky, $d$ the distance, $e$ the eccentricity. The so-called Shklovskii effect \citep{Shklovskii} arises due to the transverse motion of the binary and can be expressed as:
\begin{equation}
\PBDOTshk = \mu^2 d \frac{P_{b}}{c}.    
\end{equation}
Tidal dissipation inside the star \citep{Ogilvie2014} gives rise to : 
\begin{equation}
\dot{P_{b}}^{\rm tidal} = -\frac{27 \pi}{2} \sum\limits_{i=1,2} \frac{1}{Q^{\prime}_i}\frac{M_j}{M_i} \left(\frac{R_i}{a}\right)^{5},
\label{eq:tidal}
\end{equation}
where $j\neq i$ and $Q^{\prime}_i$ are the reduced tidal quality factors for both stars.  Here, we assume that there are equal contributions from both stars, and that the stars are not tidally synchronized.  
This is the maximum possible contribution from tidal decay because tidal decay would be suppressed if the system is synchronized.   Typical values of the reduced tidal quality factor are $Q^\prime \sim 10^{6}$ \citep{Ogilvie2014,Patra2020}, although lower values ($Q^\prime \sim 4 \times 10^{5}$) have also been inferred for short-period planets like WASP-43 b \citep{Davoudi2021}.  

When the binary is eccentric, there are also contributions to $\dot P_b$ due to the apsidal precession of the orbit. The contribution due to the GR precession \citep{Rafikov2009} is:
\begin{equation}
\PBDOTGR = \frac {36 \pi e \rm cos \omega} {(1-e^{2})^{1/2} (1+e \rm sin \omega)^{3}} \left(\frac{na}{c}\right)^{4}    
\label{eq:grprecession}
\end{equation}

Period variation caused by apsidal precession due to tidally and rotationally induced quadrupoles is :
\begin{eqnarray}
\dot{P_{b}}^{\rm quad} = \frac{4 \pi (\dot \omega^{\rm quad})^2}{n^{2}}e \rm cos \omega \frac{\left(1-e^{2}\right)^{3/2}}{\left(1+e \rm sin \omega \right)^{3}},
\label{eq:Pbdotquad}
\end{eqnarray}
where for tidally-induced quadrupole \citep{Fab2007,Philippov2013}
\begin{eqnarray}
\dot\omega_{\rm tide}^{\rm quad} = \frac{15}{8}nf_1(e)
\sum\limits_{i=1,2}k_{2,i}
\frac{M_j}{M_i}\left(\frac{R_i}{a}\right)^5,  
\label{eq:quad_tide}
\end{eqnarray}
where  $j\neq i$ and $k_{2,i}$ are the Love numbers for both stars  ($k_{2,\star}\approx 0.16$ for the Sun, \citealt{Claret2019}), and $f_1(e)=(8+12e^2+e^4)(1-e^2)^{-5}\approx 8$ for $e\to 0$. 

For the rotationally-induced quadrupole, assuming that both stellar spin axes are aligned with the orbital angular momentum axis, one has
\begin{eqnarray}
\dot\omega_{\rm rot}^{\rm quad} = nf_2(e)\sum\limits_{i=1,2}k_{2,i}\frac{M_1+M_2}{M_i}\left(\frac{R_i}{a}\right)^5\left(\frac{\Omega_i}{n}\right)^2,
\label{eq:quad_rot}
\end{eqnarray}
where $f_2(e)=(1-e^2)^{-2}\approx 1$ as $e\to 0$, and $\Omega_i$ are the stellar spin rates. 



We define the observed line-of-sight acceleration, \aobs, as
\begin{equation}
  \aobs= \frac{c \dot{P}_{b}^{\rm Obs}}{P_{b}}.
\end{equation}

\begin{figure}[ht]        
\begin{center}
\includegraphics[scale=0.45]{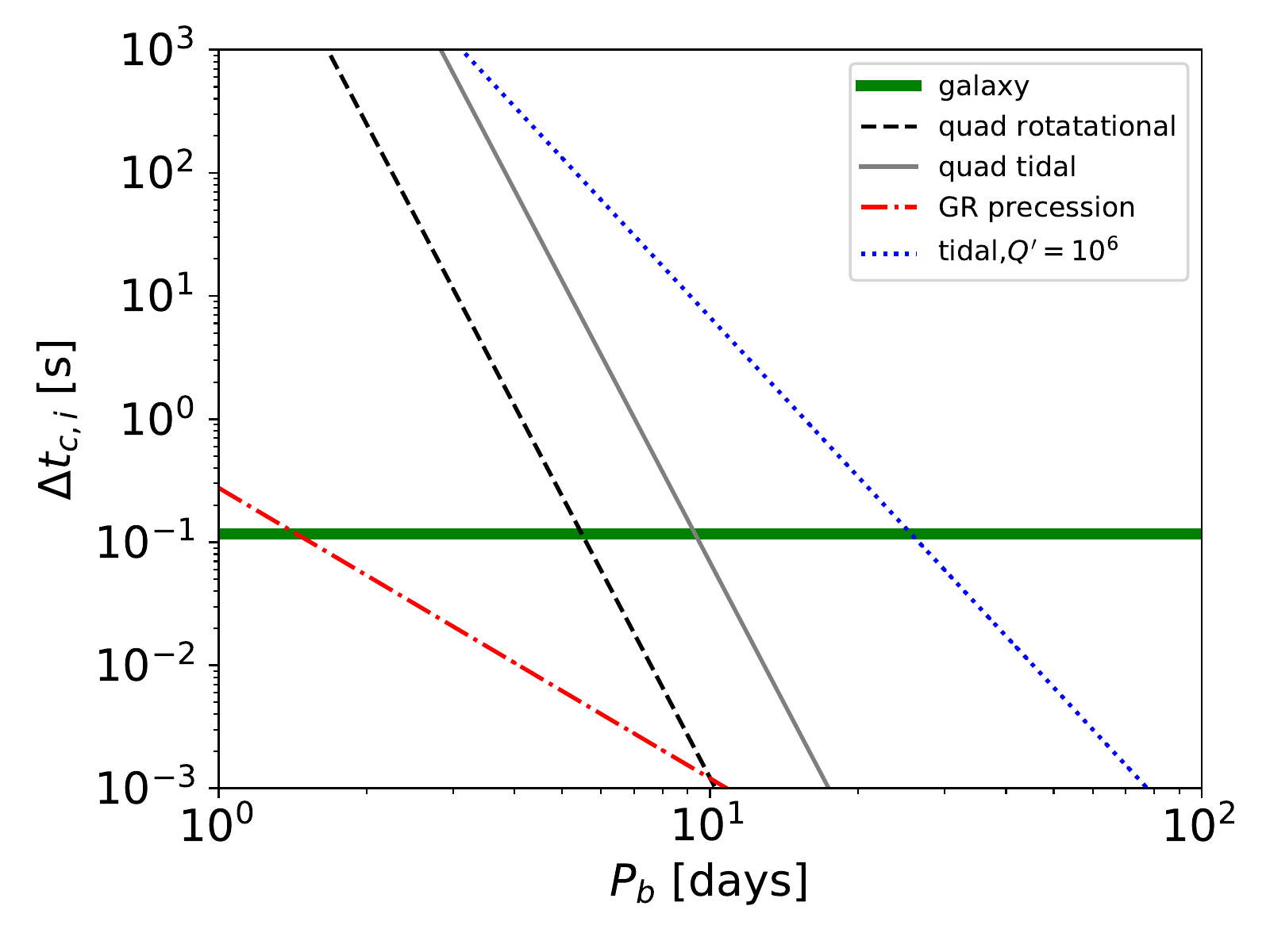}
\includegraphics[scale=0.45]{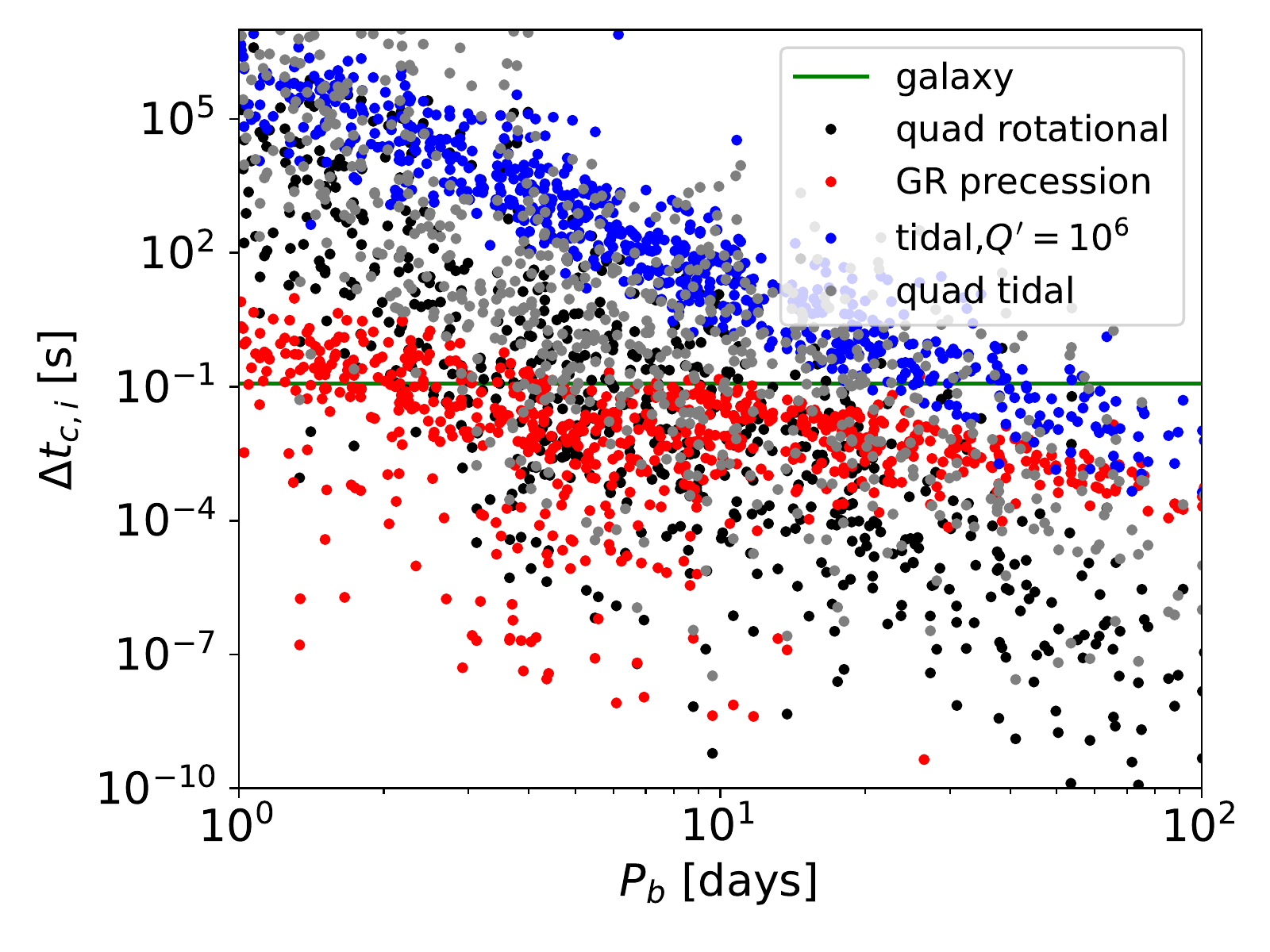}
\includegraphics[scale=0.45]{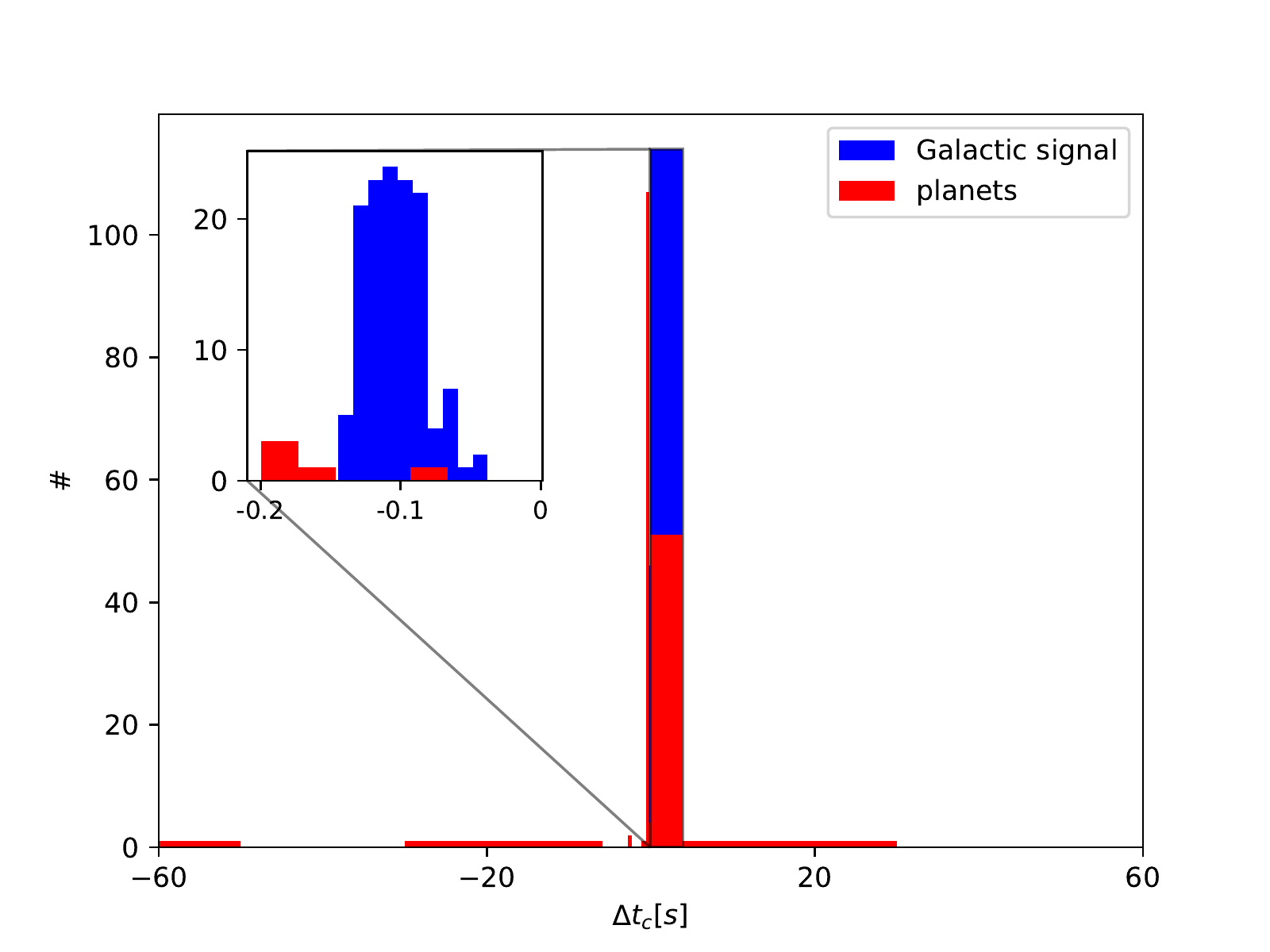}
\caption{(Top) The shift in the eclipse mid-point time $\Delta t_{c,i}$ (where $i$ refers to each individual physical acceleration mechanism) over a time-baseline of a decade due to individual acceleration mechanisms, assuming low eccentricity ($e = 0.01$).  Here, the shift in the mid-point time due to tidal decay is an upper limit, as we have assumed that the stars are not synchronized.  Assuming that the stars are synchronized, this suggests that the effective parameter space for nearly circular EBs corresponds to $P_{b} >$ 15 days, where other dynamical mechanisms lead to a $\Delta t_{c,i}$ that is smaller from that produced by the Galactic potential.  (Middle) The contributions to $\Deltci$ for the EBs analyzed by \cite{Windemuth2019}.  Assuming that the stars are tidally synchronized, there are 230 EBs for which the Galactic signal can be extracted (i.e., for set, $\Delta t_{c,\rm quad~tidal <}$ 0.1s; other contaminants to the Galactic signal are sub-dominant).  (Bottom) Histogram showing the contribution to $\Delta t_{c}$ from a realization of a synthetic population including stars that trace the Galactic signal (blue),
and circumbinary planets (red). The level of overlap is minimal, and we can expect that circumbinary planets are not a significant contaminant to the Galactic signal. \label{f:params}}
\end{center}
\end{figure}

A specific physical mechanism that induces a time-rate of the binary period is denoted $\dot{P}_{b,i}$, which then leads to a shift of $\Delta t_{c,i}$: 

\begin{equation}
\Delta t_{c,i} = \frac{1}{2} \frac{\dot{P}_{b,i}}{P_{b}} T^{2}
\end{equation}
where $T$ is the baseline of the observations covering multiple eclipses.  We take this time baseline to be a decade (roughly the elapsed time between the $\textit{Kepler}$ mission and the present).
The sum of these various mechanisms discussed above leads to a total observed $\PBDOTobs$, and a total shift in the mid-point time $\Delta t_{c}$.  To clarify which physical mechanisms are dominant in various parts of the parameter space, we plot in Figure \ref{f:params} the $\Delta t_{c,i}$ induced by an individual physical acceleration mechanism.  For systems that are circularized but not tidally synchronized, only Eq. \ref{eq:tidal} would apply, while for systems that are tidally synchronized but not circularized Eqs. \ref{eq:grprecession} and \ref{eq:Pbdotquad} would apply.

We have focused here on eclipse timing measurements of stellar EBs rather than transiting exoplanets because EBs' timing precision is generally better due to their deeper eclipses, and shorter ingress-egress durations.  The timing precision depends linearly on the transit depth, which is proportional to the square of the radius ratio of the two stars).  Additionally, the timing precision is proportional to the square root of the ingress-egress durations; these durations are also proportional to the sizes of the stars \citep{Carter2008, Winn2010}.

\subsection{The accessible parameter space for measuring Galactic accelerations with eclipse timing}

Figure \ref{f:params} (top panel) displays the contributions to the measured shift in EB mid-eclipse times $\Delta t_{c,i}$ due to the various mechanisms discussed above as a function of binary period for $e = 0.01$; for simplicity, we consider solar-mass stars with $k_{2,\star} = 0.16$. In calculating the rotationally induced quadrupole moment, we assume $\Omega_{i}/n=1$.
Here, we have assumed a tidal quality factor $Q^{\prime}=10^{6}$.
For small eccentricities, the Galactic signal is measurable for long periods ($P_{b} > 25$) days, even if the stars are not tidally synchronized (N.B. here the contribution shown from tidal decay is an upper limit).  Lowering (increasing) $Q^{\prime}$ leads to increasing (lowering) the contribution from tidal decay, such that $Q^{\prime}=10^{4}$ gives $\Deltci <$ 0.1s for $P_{b} >$ 30 days, and $Q^{\prime}=10^{7}$ gives $\Deltci <$ 0.1s for $P_{b} >$ 15 days.  Also, if the binary has a mildly eccentric orbit, the tidally induced quadrupole moment may lead to a sufficiently large $\Delta\rm t_{c,i}$ (see Equation (\ref{eq:quad_tide})) that we cannot extract the Galactic signal from an analysis of the eclipse mid-point time for shorter periods ($P_{b} < $ 20 days).
While we may expect statistically that many of the sources at low periods are circularized \citep{Justesen2021}, it is essential to calculate the orbital parameters for individual sources to explicitly check that the Galactic signal can be extracted. 



\cite{Windemuth2019} have recently presented the orbital and stellar parameters for 728 EBs observed by \textit{Kepler} for which they performed a custom extraction of the light curves.  Given their orbital and stellar parameters, we can calculate the contributions to $\Delta\rm t_{c,i}$ from the aforementioned effects for these individual systems.  We have adopted $k_{2} = 0.16$ \citep{Claret2019} in the calculation of the rotationally and tidally induced quadrupole, and have checked that for the set of stars for which these terms are sub-dominant, the sample is indeed composed of roughly solar mass stars.  The mean and standard deviation of mass of the primary star for this sample are 1.1 $M_{\odot}$ and 0.8 $M_{\odot}$ respectively, and for the secondary star are 0.98 $M_{\odot}$ and 0.5 $M_{\odot}$.
In Figure \ref{f:params} (middle) panel, we are showing the maximum possible contribution from $\dot{P_{b}}^{\rm quad}$ (by assuming synchronization), and $\dot{P_{b}}^{\rm tidal}$ (by assuming asynchronization).  Since we do not have information on the spins of the stars, we cannot say which of these possibilities is realized, but by showing both contributions we are showing a conservative estimate. 
Assuming tidal synchronization, we find that there are a large number of EBs (230) that have sufficiently low eccentricities such that the contribution to $\Delta\rm t_{c,i}$ from other physical mechanisms is lower than the Galactic acceleration, as shown in Figure \ref{f:params} (middle panel).  

Circumbinary planets may also induce a shift in the eclipse mid-point time.  Following our earlier work on examining contaminants to the Galactic signal for EPRV surveys \citep{Chakrabarti2020}, we create a synthetic population of stellar binaries and their associated circumbinary planets.  We sample from the observed demographics of planets around binary stars, which (in our current understanding) appear to be different from the single star population in several ways, as found in earlier work \citep{Armstrong2014,LiHolmanetal2016,Orosz2019,Kostov2021}.  Circumbinary planets tend to have an average distribution of periods that range from months to years, with essentially no known short-period planets.  The upper mass range of circumbinary planets is significantly less than planets orbiting single stars, and they are typically on co-planar orbits relative to the binary. 

By sampling from the observed demographics of circumbinary planets, we create a synthetic population of stars and their associated planets, and calculate the contribution to $\Deltci$ from circumbinary planets.  Here, we focus on the contribution from planets and the Galactic signal to \Deltci.  By assumption, 50 \% of the stars in our synthetic population are assigned three planetary companions, which leads to a mean number of about two planets per star.  We take the Galactic signal to be a Gaussian centered at 0.1s.  The bottom panel of Figure \ref{f:params} displays a histogram of $\Deltci$ values induced by circumbinary planets in such a synthetic population.  The p-values from the Kolmogorov-Smirnov test for the two distributions corresponding to $\pm$ 5-sigma of the mean of the Galactic signal, and circumbinary planets that overlap in this range is very small; for a typical realization the p-value is $10^{-4}$, or lower.  This indicates that these two populations are distinct, and that we can reject the null hypothesis that the signal (the measured $\Deltci$) is due to circumbinary planets. 


\section{Measuring Galactic accelerations with \textit{Kepler} EBs}

\begin{figure}[ht]        
\begin{center}
\includegraphics[scale=0.5]{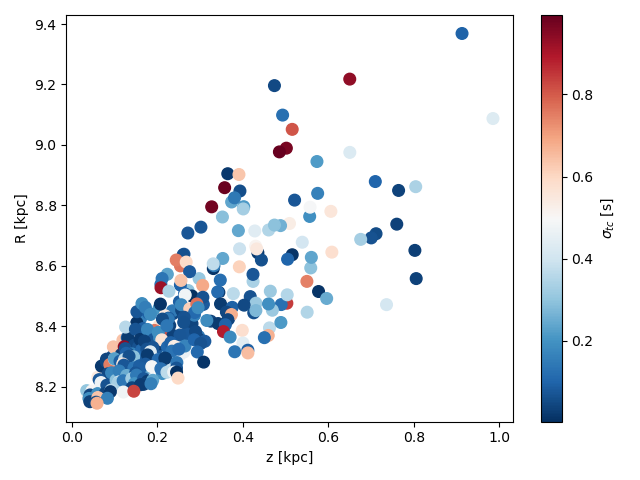}
\caption{Galactocentric R and z coordinates for the EBs from \cite{Windemuth2019} that have $\Delta t_{c,\rm quad~tidal} <$ 0.1s, and additional (70) sources from the Kepler Villanova catalog that have low eccentricities ($e < 0.01$) with timing precision better than 1s (for reference, the Sun is at R=8.1 kpc, z = 0.05 pc).  The colorbar denotes their estimated \textit{HST} mid-eclipse timing errors $\sigma_{tc}$, with bluer colors denoting better precision and redder colors denoting worse precision. We expect \textit{HST} to measure the mid-eclipse times to sub-second precision for many of these sources.  A precision of $\sim$ 0.1s allows us to measure the Galactic acceleration at $\sim$ kpc distances on a per-star basis, while lower precisions will still allow for a statistical measurement. \label{f:alos}}
\end{center}
\end{figure}

To measure Galactic accelerations over a decade time-scale it is necessary to be able to measure a shift in the eclipse mid-point time to about $\sim$ 0.1s for sources that are at $\sim$ kpc distances from the Sun.  We can expect the vertical Galactic acceleration to scale approximately linearly with vertical height; we gave a fitting formula for the vertical dependence in earlier work by analyzing pulsar timing observations \citep{Chakrabarti2021}.  This linear dependence is also expected from earlier kinematic analysis \citep{HolmbergFlynn}.  We can expect the radial component of the acceleration to scale as $V_{c}^{2}/R$, where $V_{c}$ is the circular speed and $R$ is the Galactocentric radius.  

Using the \textit{Kepler} EB Catalog (\citealt{Prsa2011,Kirk2016}), we identified detached EBs with no significant evidence of eccentricity by requiring that the catalog's morphology parameter \texttt{morph} $<0.5$ and primary/secondary eclipse separation parameter $0.49 <$ \texttt{sep} $<0.51$. For each EB, we pulled its long-cadence light curve and observable quantities (e.g. periods and eclipse durations) from the \textit{Kepler} EB Catalog, and we used the \citet{Price2014} relations to estimate the uncertainty on the mid-eclipse times. We then selected an initial sample of 70 EBs for which we estimate $\Delta t_c \leq 1$s and inspected their light curves by-eye. The \textit{Kepler} light curves of these sources typically exhibit few-hundred-ppm photometric precision.  We used the \texttt{batman} Python package \citep{Kreidberg2015} to fit transit models to these data; note that we are interested solely in the mid-eclipse timing for our purposes, and not on the accuracy of the recovered physical parameters. 

Figure \ref{f:alos} depicts the Galactocentric coordinates for these 70 EBs, which we determine from the Gaia eDR3 dataset \citep{Lindegren2021}.  We also show here the set of 230 EBs from \cite{Windemuth2019} that have a low enough tidally induced quadruopole moment that the Galactic signal should be measurable. For this set of about 300 sources, half have timing precision better than 0.5s.  Relative to our pulsar sample \citep{Chakrabarti2021}, there are a significantly larger number of EBs above the Galactic disk, and more EBs above the Galactic disk at larger radial distances from the Sun.  This suggests that EBs, in addition to pulsars, may allow for a tighter constraint on the Oort limit.  In our earlier analysis using pulsar timing, we did not see any clear patterns in the residuals of the measured line-of-sight accelerations at the pulsar locations relative to commonly used static models.  The larger number of EBs may manifest clearer residuals (that may arise due to, e.g. out-of-equilibrium effects, dark matter sub-structure, or a warp or a lopsided mass distribution) in the line-of-sight acceleration. 

\begin{figure}[h]        
\begin{center}
\includegraphics[scale=0.25]{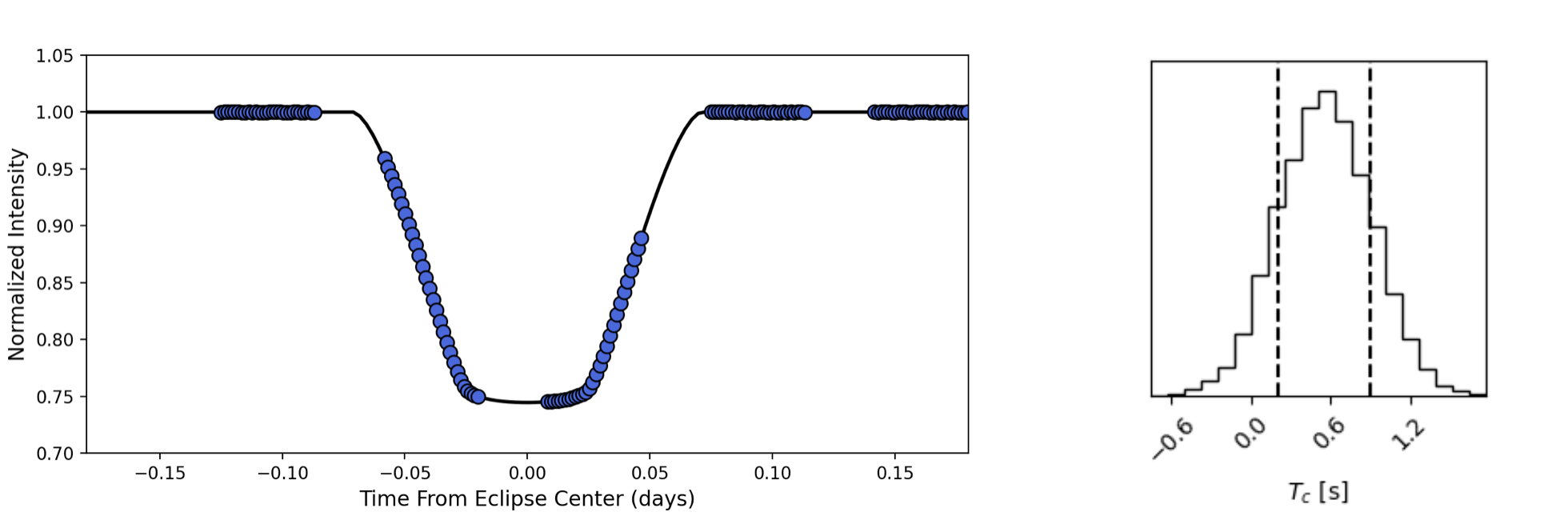}
\caption{(a) Simulated \textit{HST} WFC3 primary eclipse lightcurve of KIC 4144236 (blue points) and the best-fit transit model from \texttt{batman} (\citealt{Kreidberg2015}; black curve).  (b) Posterior distribution for the eclipse mid-point time $T_{c}$ from the transit modeling, demonstrating that a 0.1s precision on $T_{c}$ is achievable with HST.}
\label{f:lctcpost}
\end{center}
\end{figure}

To demonstrate that 0.1s eclipse timing is possible with $\textit{HST}$, we use PandExo \citep{Batalha2017} to generate a synthetic $\textit{HST}$ light curve covering the 25\% deep primary eclipse of this bright ($K = 11.9$; $J_{\rm mag}=10.93$), $P_b = 23$-day EB KIC 4144236. With 2-minute exposures, 28 exposures per $\textit{HST}$ orbit, and 6 orbits, our synthetic light curve has a typical per-point flux precision of 104ppm; by fitting a \texttt{batman} transit model to it and estimating the timing uncertainty using a MCMC analysis, we are able to determine the mid-eclipse time to 0.1s. Figure \ref{f:lctcpost} shows our synthetic light curve, best-fit transit model, and MCMC posterior for $\sigma_{t_{c}}$.

Additional variability in EB light curves can in principle reduce the accuracy of eclipse timing measurements (but not necessarily the precision). One can account for ellipsoidal variations and rotation signals from the \textit{Kepler} light curves via e.g. basis spline-fitting \citep{Vanderburg2014}. These signals' timescales are generally long relative to the eclipse durations, so they will manifest as low-order polynomial flux changes in lightcurve data. 
Follow-up spectroscopy and radial velocities can constrain EBs' eccentricities to $<1\%$ and further rule out third bodies that might contribute to ETVs.  Our analysis here, sampling from the observed demographics of circumbinary planets, suggests that dynamic effects from planets would not contaminate the Galactic signal for sufficiently large samples of acceleration measurements.

These new techniques -- pulsar timing, time-domain optical spectroscopy using extreme-precision radial velocity observations or ``optical timing'', and eclipse timing -- now enable precision measurements for ``real-time'' Galactic dynamics. 
Establishing a baseline for distant EBs will prove fruitful for upcoming space-based missions such as $\textit{Roman}$, as the signal grows quadratically with time.  Another possible way to measure accelerations is to measure changes in proper motions, but this would require a precision of $\sim$ nanoarcseconds/yr, which is not achievable by current or future astrometric missions on a single star basis.  However, \cite{Buschmann2021} have recently suggested that it may be possible to measure Galactic accelerations in aggregate by measuring changes in proper motions from current and upcoming astrometric missions.  Such a statistical measurement may be possible if systematics (spurious accelerations arising from the reference frame rotation \citep{Lindegren2018}) can be addressed.  



\section{Conclusion}
We summarize our main findings below.

$\bullet$ We show that, in principle, it is now possible to use $\textit{HST}$ (or other space-missions that can achieve comparable photometric precision) of $\textit{Kepler}$ eclipsing binaries to measure the Galactic acceleration.  By measuring an eclipse today, 
we can detect the $\sim$ 0.1s shift in the mid-point of the eclipse time due to the Galactic gravitational potential. 

$\bullet$ We have analyzed contributions to the period drift rate from sources other than the Galactic potential, including the relativistic precession of an eccentric orbit, tidal decay, and the rotationally and tidally induced quadrupole.  Our approach here in modeling these contributions is necessarily approximate due to the uncertainties in modeling tidal decay, and our calculation of the tidal decay contribution is an upper limit assuming the stars are not tidally synchronized.  For low periods, there can be significant contributions from the GR precession and the tidally and rotationally induced quadrupole even for low eccentricities.

$\bullet$ We calculate the contributions to the change in the eclipse mid-point time due to these dynamical effects for the sample of EBs analyzed earlier by \cite{Windemuth2019}, using their orbital and stellar parameters.  For this dataset, assuming that the stars are tidally synchronized, there are 230 sources with $\Deltci < $ the Galactic signal, such that the Galactic acceleration is indeed measurable.

$\bullet$ We create a synthetic population of circumbinary planets by sampling from their observed demographics and calculate their contribution to $\Deltci$.  We find that there is sufficiently small overlap with the Galactic signal such that circumbinary planets are not a significant contaminant to the Galactic signal. 

$\bullet$ Using analytic relations for the timing precision, we have found an additional 70 sources from the \textit{Kepler} EB Catalog that have timing precision better than 1s.  These EBs populate a different region of physical parameter space than the pulsars that we earlier analyzed; together with constraints from pulsar timing, they can yield a better measurement of the Oort limit.  We may be able to leverage the size of the larger EB sample (compared to the pulsar sample) to place constraints on non-equilibrium effects and/or dark matter sub-structure by analyzing residuals in the line-of-sight acceleration relative to static models.

$\bullet$ We used the EB KIC 4144236 ($K = 11.9; J_{\rm mag} = 10.93$) as a worked example, analyzing a simulated \textit{HST} WFC3 lightcurve ($\sim 100$ppm flux precision in 2-min exposures) of its 25\% deep primary eclipse. The light curve has a photometric precision of 104ppm, and we are able to constrain the mid-eclipse time to 0.1s.

$\bullet$ Unlike kinematic methods, the signal from direct acceleration methods like this one grows with time, and in this case, quadratically with time.  Thus, the Galactic signal from EBs for upcoming $\textit{Roman}$ observations will be substantially larger than what it is today (when the measurement has just become possible), given the baseline established a decade ago by $\textit{Kepler}$.

\bigskip
\bigskip 

\acknowledgements

The Center for Exoplanets and Habitable Worlds and the Penn State Extraterrestrial Intelligence Center are
supported by the Pennsylvania State University and the Eberly College of Science.  

SC gratefully acknowledges support from NSF AAG 2009574, and hospitality provided by the CCA at the Flatiron Institute. RRR is supported by STFC grant ST/T00049X/1. We thank N. Sehgal for coining the word "optical timing".  We thank J. Winn, S. Tremaine,  R. Dawson, G. Ogilvie, A. Prsa, and J. Pepper for helpful discussions.

This research has made use of 
NASA's Astrophysics Data System Bibliographic Services.  This work has made use of data from the European Space Agency (ESA) mission
{\it Gaia} (\url{https://www.cosmos.esa.int/gaia}), processed by the {\it Gaia}
Data Processing and Analysis Consortium (DPAC,
\url{https://www.cosmos.esa.int/web/gaia/dpac/consortium}). Funding for the DPAC
has been provided by national institutions, in particular the institutions
participating in the {\it Gaia} Multilateral Agreement.

\bibliographystyle{aasjournal}
\bibliography{bibl}

\end{document}